\documentclass[referee]{aa} 

\usepackage{graphicx}
\usepackage{txfonts}
\begin{document}

   \title{Signatures of coronal hole substructure in the solar wind: combined Solar Orbiter remote sensing and in situ measurements}


   \author{T. S. Horbury\inst{1}\thanks{Corresponding author: Tim Horbury \email{t.horbury$@$imperial.ac.uk}}
\and          
R. Laker\inst{1}
\and
L. Rodriguez\inst{8}
\and
K. Steinvall\inst{10}
\and
M. Maksimovic\inst{11}
\and
S, Livi\inst{12}
\and
D. Berghmans\inst{8}
\and
F. Auchere\inst{9}
\and
A. N. Zhukov\inst{8,15}
\and
Yu. V. Khotyaintsev\inst{10}
\and
L. Woodham\inst{1}
\and    
L. Matteini\inst{1}
\and    
J. Stawarz\inst{1}
\and    
T. Woolley\inst{1}
\and    
S. D. Bale\inst{2,3}
\and 
A. Rouillard\inst{6}
\and 
H. O'Brien\inst{1}
\and
V. Evans\inst{1}
\and
V. Angelini\inst{1}
\and
C. Owen\inst{13}
\and
S. K. Solanki\inst{7,15}
\and
B. Nicula\inst{8}
\and
D. M\"uller\inst{14}
\and
I. Zouganelis\inst{14}
}

   \institute{Imperial College London, South Kensington Campus, London SW7 2AZ, UK
\and   
  Physics Department, University of California, Berkeley, CA 94720-7300, USA
\and   
   Space Sciences Laboratory, University of California, Berkeley, CA 94720-7450, USA
\and
   Climate and Space Sciences and Engineering, University of Michigan, Ann Arbor, MI 48109, USA
\and
   Smithsonian Astrophysical Observatory, Cambridge, MA 02138 USA.
\and
    Institut de Recherche en Astrophysique et Planétologie (IRAP), CNRS, University of Toulouse, Toulouse, France
    \and
    Max-Planck-Institut f\"ur Sonnensystemforschung, 37077, G\"ottingen, Germany
\and
Solar-Terrestrial Centre of Excellence - SIDC, Royal Observatory of Belgium, Brussels, Belgium
\and
Institut d'Astrophysique Spatiale, Paris
\and
   Space and Plasma Physics, Department of Physics and Astronomy, Uppsala University, Uppsala 75120, Sweden
   \and
   LESIA, Observatoire de Paris, Meudon, Paris
   \and
   Southwest Research Institute, San Antonio, Texas
   \and
   Mullard Space Science Laboratory, University College London
   \and
   European Space Agency
   \and
   Skobeltsyn Institute of Nuclear Physics, Moscow State University, Moscow, Russia
   \and
   School of Space Research, Kyung Hee University, Yongin, 446-701, Republic of Korea
             }

   \date{Received XXXX; accepted XXXX}

 
  \abstract
   {The Sun's complex corona is the source of the solar wind and interplanetary magnetic field. While the large scale morphology is well understood, the impact of variations in coronal properties on the scale of a few degrees on properties of the interplanetary medium is not known. Solar Orbiter, carrying both remote sensing and in situ instruments into the inner solar system, is intended to make these connections better than ever before.}
   {We combine remote sensing and in situ measurements from Solar Orbiter's first perihelion at 0.5~AU to study the fine scale structure of the solar wind from the equatorward edge of a polar coronal hole with the aim of identifying characteristics of the corona which can explain the in situ variations.}
   {We use in situ measurements of the magnetic field, density and solar wind speed to identify structures on scales of hours at the spacecraft. Using Potential Field Source Surface mapping we estimate the source locations of the measured solar wind as a function of time and use EUI images to characterise these solar sources. 
   }
   {We identify small scale stream interactions in the solar wind with compressed magnetic field and density along with speed variations which are associated with corrugations in the edge of the coronal hole on scales of several degrees, demonstrating that fine scale coronal structure can directly influence solar wind properties and drive variations within individual streams.}
   {This early analysis already demonstrates the power of Solar Orbiter's combined remote sensing and in situ payload and shows that with future, closer perihelia it will be possible dramatically to  improve our knowledge of the coronal sources of fine scale solar wind structure, which is important both for understanding the phenomena driving the solar wind and predicting its impacts at the Earth and elsewhere.}


\titlerunning{Solar wind signatures of coronal substructure}
\authorrunning{Horbury et al.}

   \maketitle
%

\section{Introduction}

The large scale structure of the solar wind and magnetic field and their variation over the solar cycle are well established \citep[e.g.][]{RN1814,RN1894}. Fast, highly Alfv\'enic solar wind flows from magnetically open ``coronal holes,'' which cover the polar regions near solar minimum when the Sun's global field is approximately dipolar, but can occur at any latitude near solar maximum when the field is far more complex. The source of slower wind is less clear, coming either from the edges of coronal holes or active regions. Transient events, from large scale coronal mass ejections to smaller quasi-periodic ejecta are also present. Solar rotation means that steady solar wind streams with longitudinal extents of tens of degrees result in variations at spacecraft near 1~au of one to several days and interactions between streams of different speeds, at least away from the Sun's poles, result in corotating interaction regions \citep[CIRs:][]{RN1886, RN1888} which at spacecraft have a characteristic sequence of slow wind, followed by a speed increase associated with compressed field and plasma, and then a higher speed following stream. 

Within this large scale picture, however, there is much that is unclear \citep{RN1891}. In particular, there is ubiquitous solar wind structure on scales of hours to a day or so, whose origin is not known. Transient events within the streamer belt are common \citep[e.g.][]{RN1889} but even within coronal hole flows there is often considerable structure, at least near the boundary. 

Our knowledge of the structure and dynamics of the near-Sun solar wind has been dominated by the extensive and detailed analysis of data from the twin Helios probes which travelled to 0.3~au in the 1970's and 1980's, much of which is comprehensively summarised in \cite{1990pihl.book...99S}. Solar wind velocity variations at 0.3~au were shown to have sharp longitudinal boundaries \citep{RN1892}, with typical scales of around $5^{\circ}$, and even smaller latitudinal scales of $1-2^{\circ}$. Interactions between plasma parcels of different speeds, as they travel anti-sunward, means that by 1~au such structure is considerably processed, and indeed linking solar wind features to solar structures in scales of a few degrees is very challenging. The closer that measurements are made to the Sun, the easier it is to make this link.

One major limitation on the ability to link solar wind structures at Helios to their sources was the lack of high quality remote sensing measurements. Since the Helios era, remote sensing instrumentation has improved markedly and we now have access to routine, high quality EUV imaging of the Sun, as well as magnetograms which make it possible to determine the large scale solar magnetic field structure which controls the global solar wind configuration. Most recently, Parker Solar Probe has explored under 20 solar radii ($\mathrm{R_{S}}$) and revealed a wealth of fine scale solar wind structure. Parker data have been linked to solar coronal and magnetic field structure \citep{RN1878, RN1877, RN1876} but to date not at very small scales. Without on-disc telescopes, Parker perihelia often occur without simultaneous imaging, limiting the opportunities for making these connections.

Solar Orbiter \citep{RN1884} carries a unique payload of  remote sensing and in situ instruments with the goal of determining how processes on the Sun affect those in interplanetary space. By travelling close to the Sun it can measure plasma and fields in a relatively young state before stream-stream interactions affect them; and it can simultaneously characterise the source regions with its comprehensive suite of telescopes. Crucially, by co-locating the two sets of measurements, the solar source regions can always be imaged, regardless of the spacecraft location. Here, we use the data from Orbiter's first perihelion to investigate the connections between solar, and solar wind, structures.

\section{Solar Orbiter's first perihelion}
Between 17 and 26 June 2020, during the first check-out of the remote sensing instruments, Solar Orbiter moved between 0.51 and 0.54~au, and $6-5^{\circ}N$. In situ magnetic field measurements \citep{RN1860} are available throughout but only HIS sensor \citep{RN1863} estimates of the alpha particle bulk speed are available from 22-26 June. Estimates of the solar wind deHoffmann-Teller (dHT) frame radial wind speed have therefore been made by combining the RPW \citep{RN1861} electric field and MAG data, as described by \cite{Steinvall21}. We use RPW electron number density estimates.

\begin{figure}
   \centering
   \includegraphics[width=9cm]{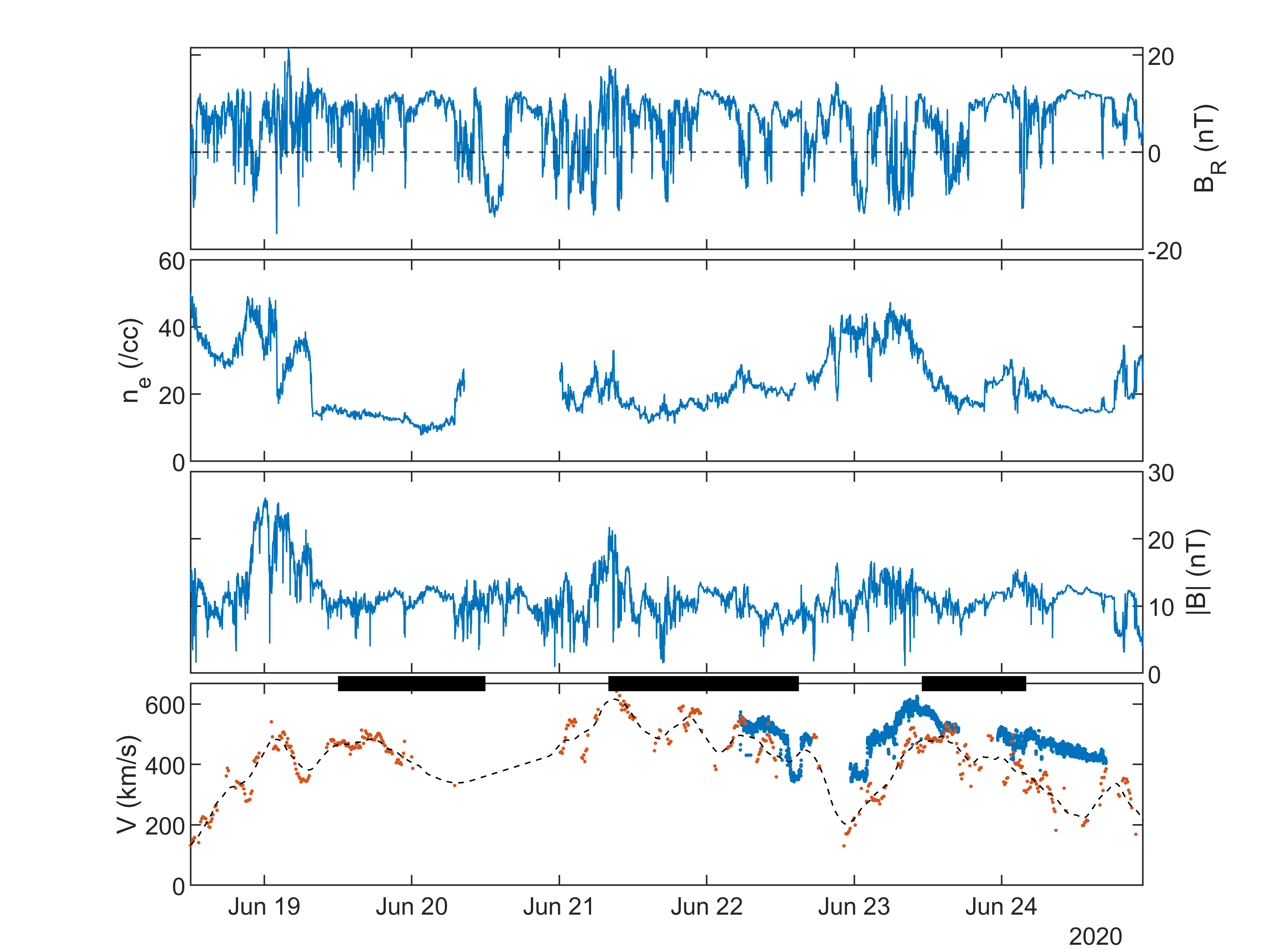}
   \caption{In situ Solar Orbiter observations. Panels are from top to bottom: radial magnetic field component; electron number density; magnetic field magnitude; estimated solar wind speed (blue: HIS $He^{++}$, red: RPW-MAG deHoffmann-Teller (dHT) speed, dashed: 6 hour smoothed dHT. Horizontal bars correspond to the same periods as Figure \ref{fig:MappedLongitudes}}
              \label{fig:OrbiterTime}%
    \end{figure}

The in situ data are shown in Figure \ref{fig:OrbiterTime}. The magnetic polarity throughout this interval was positive, consistent with connection to wind from the Sun's Northern magnetic hemisphere, albeit with large amplitude excursions largely due to switchbacks. Variations in density and field magnitude were generally positively correlated, consistent with compression and rarefaction due to interactions between plasma streams at different speeds, rather than pressure balanced structures. The solar wind speed was variable. Here, HIS alpha speeds and dHT frame estimates (red dots) are both shown, with broad agreement when both are present but some systematic differences. We use dHT estimates in preference to HIS due to their wide coverage -- the dashed line is a smoothed versions of this data. These should still be treated with caution: the very lowest speed estimates are probably anomalously low, and we only have a radial speed estimate, so must neglect potential non-radial flows. Nevertheless, some consistent structure is visible and speed increases of order $100~\mathrm{km/s}$ are typically associated with periods of enhanced density and field magnitude, on spacecraft scales of hours to around a day. This again is consistent with compression due to flows with different speeds -- essentially, miniature versions of stream interaction regions \citep[e.g.][]{RN1886} which can last several days at $1~\mathrm{au}$ and are typically associated with interactions between streams with larger speed differences.

\section{Identifying the sources of fine scale solar wind structure}
The speed, density and field variations visible in Figure \ref{fig:OrbiterTime} all occur within one solar wind stream, of a single polarity. Such small scale features are common in the solar wind, particularly near the streamer belt and in slow wind, but it is normally effectively impossible to identify a possible solar origin. Solar Orbiter's proximity to the Sun and remote sensing instruments, combined with magnetic field mapping into the corona, makes such identification possible for the first time. 

\begin{figure}
   \centering
   \includegraphics[width=9cm]{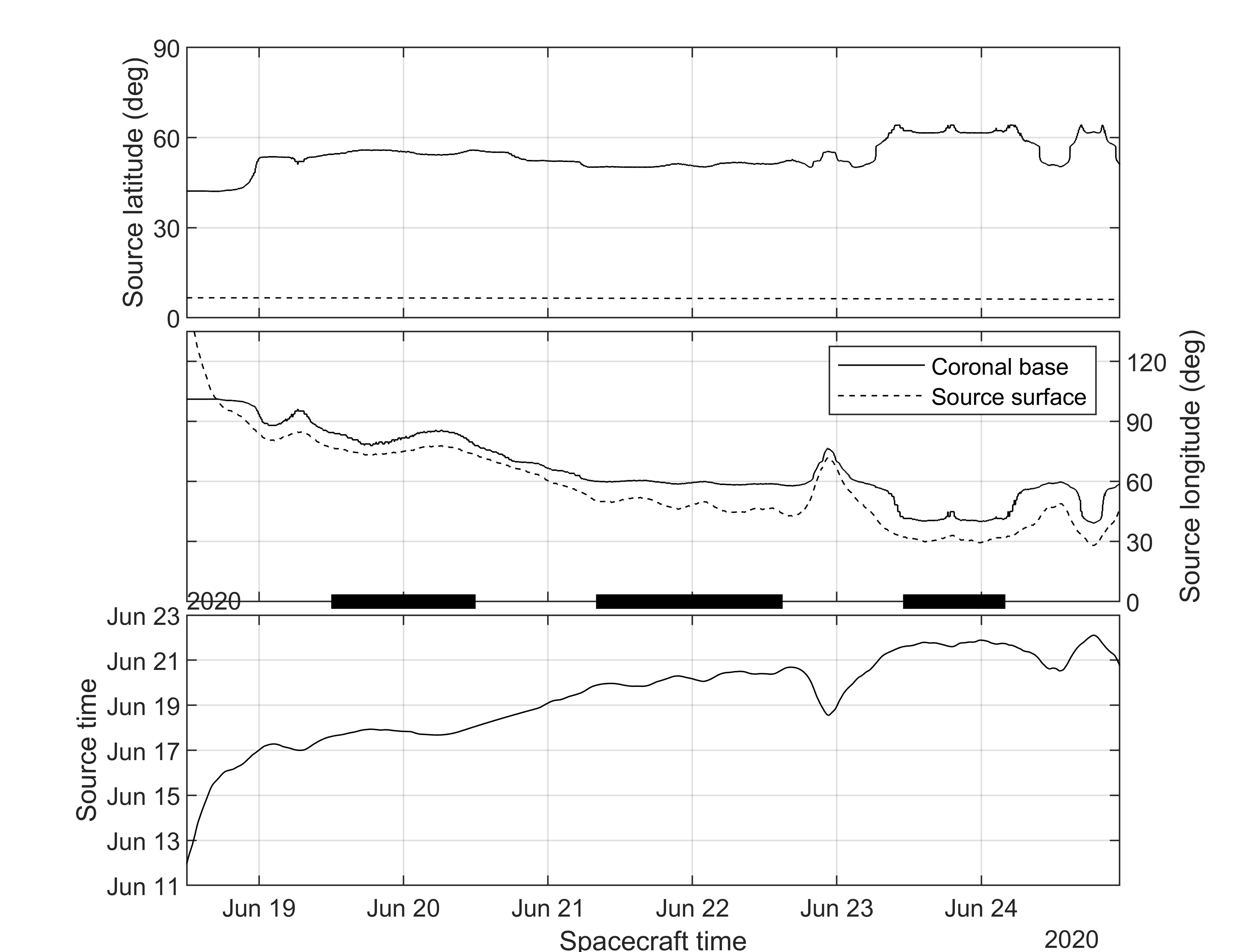}
   \caption{Latitude (top) and Carrington longitude (middle) at the solar surface (solid line) and source surface (dashed line), as a function of arrival time at the spacecraft, based on ballistic mapping using the estimated dHT frame solar wind speed. Bottom panel shows the estimated time at which the plasma left the source surface. Horizontal bars indicate times when the solar wind source remained at one location for an extended period.}
              \label{fig:MappedLongitudes}%
    \end{figure}

To make this connection, we must know the origin in the corona of the solar wind which arrives at the spacecraft. This involves two steps. The first is a ``ballistic mapping'' of the wind from the spacecraft, to a spherical source surface which we here take to be $2.5~\mathrm{R_S}$. This well-established method \citep[e.g.][]{1990pihl.book...99S, RN1878} requires an estimate of the solar wind speed and we have used the smoothed dHT frame estimate shown as a dashed line in Figure \ref{fig:OrbiterTime}. This takes into account the varying position of the spacecraft and the time it takes the wind to propagate to the spacecraft, typically slightly under 2 days for $500~\mathrm{km/s}$ wind. We stress that this  estimate assumes a constant (radial) wind flow, with for example no in-flight variations due to stream-stream interactions. In some ways this is more accurate closer to the Sun, where such interactions have less time to develop, but in turn other effects such as acceleration of the wind with radius become more significant; in future, a combination of better in situ measurements with more sophisticated models of solar wind acceleration and interactions will produce more reliable mappings.  The source surface locations of the wind, as a function of time of arrival at Solar Orbiter, are shown in Figure \ref{fig:MappedLongitudes} as dashed lines, along with the assumed time it left the corona; note that the very slow wind measured at the beginning of the 23rd of June has what is probably an unrealistically early release time due to the extremely low estimated speed and hence probably an unrealistically high source longitude.

We have attempted to assess the quality of our mappings by changing the estimated solar wind speeds, and spacecraft locations, and find that the overall picture is unchanged for $\approx100~km/s$ speed variations and spacecraft latitudes of $1-2^{\circ}$ -- the coronal magnetic configuration at this time was robust to such variations.

The second step uses a magnetogram and a simple Potential Field Source Surface (PFSS) model \citep{1969SoPh....6..442S} to follow magnetic field lines from the source surface to the base of the corona. PHI \citep{RN1881} was operational during this time, but at this stage magnetograms have not been merged with Earth-based data to provide a product which allows the mapping of solar wind directly to its source. As a result, we use ADAPT data \citep{RN1885} based on GONG magnetograms to map between the source surface and the corona: given the position of Orbiter, the GONG data used was just a few days old but evolution of the solar field could occur over this time which cannot be captured using these data. Far more sophisticated approaches are possible, but as we will see this is adequate for our purposes. The mapped source locations in the lower corona are shown in Figure \ref{fig:MappedLongitudes} as solid lines, where the differences from the source surface locations are due to the non-radial field lines between the surface and $2.5~R_{S}$.

At the end of this procedure, we have, for every moment in time at Solar Orbiter, an estimate of the origin in the lower corona of the plasma and magnetic field that the spacecraft encountered. A striking feature of this mapping is that while the source surface longitudes are rather variable (although generally decreasing with time due to solar rotation), those in the lower corona are characterised by periods of relatively constant value, interspersed by sharp jumps -- this is also clear in the origin latitudes. Three such periods are marked on Figures \ref{fig:OrbiterTime} and \ref{fig:MappedLongitudes} with horizontal bars: 19-20~June  when the coronal source location remained at around ($80^{\circ},55^{\circ}N$), 21-23~June at ($60^{\circ},50^{\circ}N$) and 23-24~June at ($40^{\circ},60^{\circ}N$). These correspond to times where the source region of the wind remained relatively constant. The causes of these stable source regions are twofold: first, that speed variations can result in ``dwells" \citep{RN1883} as commonly seen in the trailing edges of high speed streams, where several days of wind can map to the same source surface longitude; and second, that the field mapping from the source surface to the lower corona takes quite large longitudinal ranges at $2.5~R_{S}$ onto small regions within coronal holes. There are a few times when the source surface longitude increases slightly which probably reflect errors in the speed estimates, non-radial flows or interactions between plasma parcels as they travelled from the Sun to the spacecraft.

\begin{figure*}
   \centering
   \includegraphics[trim={0cm 2.3cm 0 3cm},clip,width=12cm]{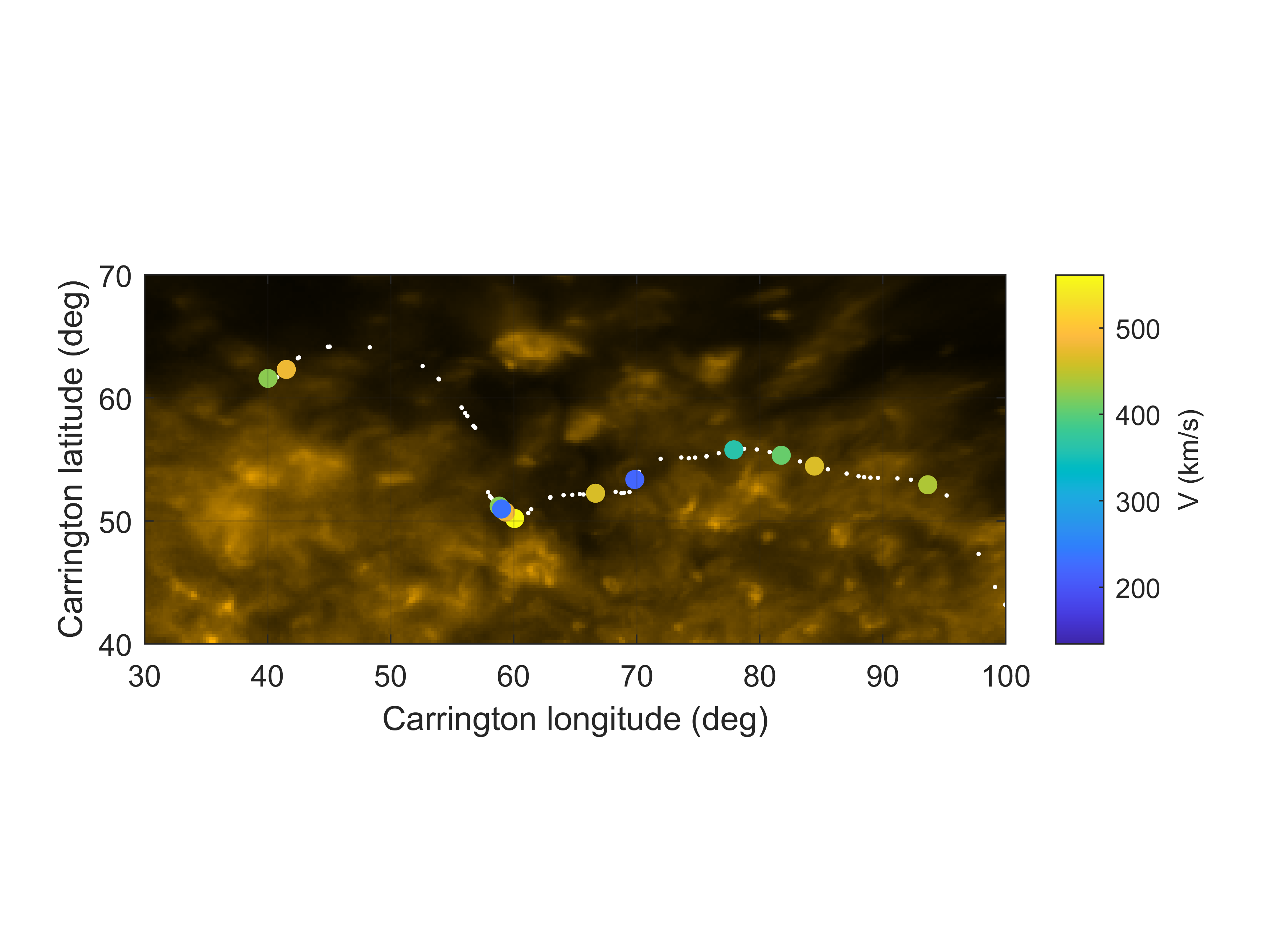}
   \caption{Solar wind source locations in the lower corona. Underlying image is a 174~\AA ~EUI composite taken on 18-20 June 2020. Small white dots show the estimated source location of the plasma measured at Solar Orbiter, at 1 hour cadence. Coloured dots show the estimated in situ solar wind speed, with a cadence of 12 hours. 
   }
              \label{fig:EUIPlot}%
    \end{figure*}

We use images from EUI \citep{RN1880} to provide direct imaging of the source regions of the solar wind. A composite EUI image of the corona at 174~\AA ~is shown in Figure \ref{fig:EUIPlot}, built up from 30 images taken from 18-20~June centred on the nadir central meridian.  Overlaid are the mapped source coronal locations of the measured wind: white dots are shown for each hour at Solar Orbiter. Larger dots every 12 hours are coloured by the measured wind speed. The Orbiter track was right to left, with the source region of the 21-23~June wind clear at ($60^{\circ},50^{\circ}N$) as a concentrated clump of dots. This was near the edge of the Northern polar coronal hole, in an equatorward extension. The source of the 23~June wind at ($40^{\circ},60^{\circ}N$) is also clear as the equatorward edge of the hole, but significantly farther North. Between these, the mapping moves rapidly (clear in Figure \ref{fig:MappedLongitudes}) and it is between these two stable regions that it suggests the slower wind, at the beginning of the 23rd of June, originated.

\section{Discussion}

We have seen that, unsurprisingly, combined ballistic and PFSS mapping predicts that the plasma and field measured by Solar Orbiter just North of the equator originated from near the equatorial edge of the polar coronal hole. The latitudinal variations in this edge explain the variations in source latitudes in Figure \ref{fig:MappedLongitudes}. Such ripples in the edges of polar coronal holes, on scales of around $10-20^{\circ}$  in longitude, are common near solar minimum. The question naturally arises whether any solar wind features can be explained by this coronal hole structuring.

The period on 21-21 June when the plasma measured by Orbiter appears to have originated from a relatively small region of the coronal hole and small speed variations are present. Patches of switchbacks \citep[e.g.][]{RN1854, RN1848} are prevalent, interspersed with periods when the field is more radial and quieter. This alternation of patches and quiet radial regions is commonly seen in Parker data closer to the Sun and has more recently been interpreted (Bale et al., in preparation) as being due to passing into and out of flows from plumes \citep[see, e.g.][]{RN1864} -- whatever the cause, it seems to be a signature of spatial structure in the coronal hole flow. The radial periods for 19-20 June, 21-22 June, and 23-24 June in Figure \ref{fig:OrbiterTime} are many hours long, unusual in the solar wind. These periods, when the effective point on the Sun from which the wind originated was moving slowly, might therefore reflect the spatial scale of these structures. In other words, the duration of these regions in the Orbiter time series might be long due to the fact that the spacecraft remained connected to the same source region for much longer than would be expected for a pure radial field expansion and meant that the time taken to cross over a flow from a given structure was unusually long. \cite{RN926} have previously noted that the longitudinal scales of structures in the solar wind need not correspond directly to those on the Sun due to over-expansion of the outflow at low latitudes.

We now consider the feature on the 22-23 June, with slow wind at the end of the 22nd. The dHT frame speed estimate here seems extremely low, but it is nevertheless clear from both this and the HIS data that there was a speed decrease, followed by an increase, during which the field and density were enhanced. This is similar to a corotating stream interaction region, just on a smaller scale, but with only radial speed data we cannot confirm the expected flow deflection \citep{RN1888}. The ultimate cause of the structure is slow wind flowing between fast wind from deeper within the coronal hole. As the source footpoints moved rapidly during this time, Orbiter was likely to have been connected to either the edge of the coronal hole, from where we would expect the wind to be slower than deeper into the hole due to flux tube expansion \citep[e.g.][]{RN1890}, or perhaps into the streamer belt \citep[e.g.][]{RN1887}. In either case, the configuration of the magnetic field resulted in slower wind surrounded by fast at Orbiter's latitude, and hence the stream interaction feature.

It is also possible that the slow wind at the end of the 22nd was a transient, rather than corotating, feature. A small CME erupted on the 21st of June, but significantly Eastward of the Orbiter footpoint locations (around 345, 40N), and it seems unlikely that this was related to the observed solar wind feature. Several weak brightenings occurred within the coronal hole around the time of origin of the slow wind, as is usual within coronal holes, but no clear precursor. The speed and density profile of the event is also more consistent with compression than a slow transient. We conclude that the observed structure was probably not caused by a transient, but was more likely to be corotating.

There seems to have been a structure early on the 21 June similar to that on the 23rd, with enhanced field and density consistent with compression and this is also associated with the time when the mapping jumps rapidly in longitude, but a RPW data gap means we do not have a speed estimate to be confident that this was associated with preceding slower speed plasma.

We conclude that the corrugated edge of the Northern polar coronal hole and complex magnetic field mapping from the surface into interplanetary space resulted in the solar wind feature observed on the 23rd of June, with speed variations and associated field and density compressions. Even such relatively crude mapping as was performed here, combined with Orbiter's proximity to the Sun, make it possible to estimate the source location of the observed plasma which appears to correspond well to the observed in situ structures during this time. 

Many of the features here are unique to solar minimum conditions, with polar coronal holes without large equatorward extensions. When coronal holes extend to lower latitudes, we might expect much stronger speed variations at a given latitude and hence well developed CIRs with less of the fine scale structure that Orbiter measured during the period discussed here. As solar activity increases, Orbiter's unique combination of remote sensing and in situ instruments will enter nominal science operations, providing coordinated measurements within the inner heliosphere: combined with near-Earth remote sensing and other in situ measurements including those from Parker Solar Probe, they should make it possible to determine the impact in the solar wind of progressively smaller scale variability at the solar surface, and hence make better predictions of conditions at the Earth, the Moon, Mars and elsewhere in the Solar System.

\begin{acknowledgements}
      Tim Horbury and Lloyd Woodham were supported by STFC grant ST/S000364/1, Thomas Woolley by ST/N504336/1 and Ronan Laker by an Imperial College President's scholarship. Solar Orbiter magnetometer operations are funded by the UK Space Agency (grant ST/T001062/1). We thank CNES, CNRS, the Paris Observatory, The Swedish National Space Agency, ESA-PRODEX and all the involved institutes for their funding. Solar Orbiter Solar Wind Analyser (SWA) data are derived from scientific sensors which have been designed and created, and are operated under funding provided in numerous contracts from the UK Space Agency (UKSA), the UK Science and Technology Facilities Council (STFC), the Agenzia Spaziale Italiana (ASI), the Centre National d’Etudes Spatiales (CNES, France), the Centre National de la Recherche Scientifique (CNRS, France), the Czech contribution to the ESA PRODEX programme and NASA. Solar Orbiter SWA work at UCL/MSSL is currently funded under STFC grants ST/T001356/1 and ST/S000240/1. Solar Orbiter data are available from the Solar Orbiter Archive at \url{http://soar.esac.esa.int/soar/}. EUI data in this study are part of data release 1, \url{	https://doi.org/10.24414/wvj6-nm32}. The EUI instrument was built by CSL, IAS, MPS, MSSL/UCL, PMOD/WRC, ROB, LCF/IO with funding from the Belgian Federal Science Policy Office (BELPSO); the Centre National d’Etudes Spatiales (CNES); the UK Space Agency (UKSA); the Bundesministerium für Wirtschaft und Energie (BMWi) through the Deutsches Zentrum für Luft- und Raumfahrt (DLR); and the Swiss Space Office (SSO). L. R. and A.N.Z. thank the European Space Agency (ESA) and the Belgian Federal Science Policy Office (BELSPO) for their support in the framework of the PRODEX Programme. PFSS modelling used the pfsspy package \citep{Stansby2020}. JHelioviewer \citep{RN1893} is part of the ESA/NASA Helioviewer Project. Solar Orbiter is a space mission of international collaboration between ESA and NASA, operated by ESA.   
\end{acknowledgements}

  \bibliographystyle{aa} 

\begin{thebibliography}{29}
\expandafter\ifx\csname natexlab\endcsname\relax\def\natexlab#1{#1}\fi

\bibitem[{Allen {et~al.}(2020)Allen, Lario, Odstrcil, Ho, Jian, Cohen, Badman,
  Jones, Arge, Mays, Mason, Bale, Bonnell, Case, Christian, de~Wit, Goetz,
  Harvey, Henney, Hill, Kasper, Korreck, Larson, Livi, MacDowall, Malaspina,
  McComas, McNutt, Mitchell, Pulupa, Raouafi, Schwadron, Stevens, Whittlesey,
  \& Wiedenbeck}]{RN1876}
Allen, R.~C., Lario, D., Odstrcil, D., {et~al.} 2020, Astrophysical Journal
  Supplement Series, 246

\bibitem[{Arge {et~al.}(2010)Arge, Henney, Koller, Compeau, Young, MacKenzie,
  Fay, \& Harvey}]{RN1885}
Arge, C.~N., Henney, C.~J., Koller, J., {et~al.} 2010, in AIP Conference
  Proceedings, Vol. 1216, 12th International Solar Wind Conference (MELVILLE:
  Amer Inst Physics), 343--+

\bibitem[{Badman {et~al.}(2020)Badman, Bale, Oliveros, Panasenco, Velli,
  Stansby, Buitrago-Casas, Reville, Bonnell, Case, de~Wit, Goetz, Harvey,
  Kasper, Korreck, Larson, Livi, MacDowall, Malaspina, Pulupa, Stevens, \&
  Whittlesey}]{RN1878}
Badman, S.~T., Bale, S.~D., Oliveros, J.~C., {et~al.} 2020, Astrophysical
  Journal Supplement Series, 246

\bibitem[{Bale {et~al.}(2019)Bale, Badman, Bonnell, Bowen, Burgess, Case,
  Cattell, Chandran, Chaston, Chen, Drake, De~Wit, Eastwood, Ergun, Farrell,
  Fong, Goetz, Goldstein, Goodrich, Harvey, Horbury, Howes, Kasper, Kellogg,
  Klimchuk, Korreck, Krasnoselskikh, Krucker, Laker, Larson, MacDowall,
  Maksimovic, Malaspina, Martinez-Oliveros, McComas, Meyer-Vernet, Moncuquet,
  Mozer, Phan, Pulupa, Raouafi, Salem, Stansby, Stevens, Szabo, Velli, Woolley,
  \& Wygant}]{RN1848}
Bale, S.~D., Badman, S.~T., Bonnell, J.~W., {et~al.} 2019, Nature, 576, 237

\bibitem[{Cranmer {et~al.}(2017)Cranmer, Gibson, \& Riley}]{RN1814}
Cranmer, S.~R., Gibson, S.~E., \& Riley, P. 2017, Space Science Reviews, 212,
  1345

\bibitem[{Di~Matteo {et~al.}(2019)Di~Matteo, Viall, Kepko, Wallace, Arge, \&
  MacNeice}]{RN1889}
Di~Matteo, S., Viall, N.~M., Kepko, L., {et~al.} 2019, Journal of Geophysical
  Research-Space Physics, 124, 837

\bibitem[{Gosling \& Pizzo(1999)}]{RN1888}
Gosling, J.~T. \& Pizzo, V.~J. 1999, Space Science Reviews, 89, 21

\bibitem[{Higginson \& Lynch(2018)}]{RN1887}
Higginson, A.~K. \& Lynch, B.~J. 2018, Astrophysical Journal, 859, 13

\bibitem[{Horbury {et~al.}(2020)Horbury, O'Brien, Blazquez, Bendyk, Brown,
  Hudson, Evans, Oddy, Carr, Beek, Cupido, Bhattacharya, Dominguez, Matthews,
  Myklebust, Whiteside, Bale, Baumjohann, Burgess, Carbone, Cargill, Eastwood,
  Erdos, Fletcher, Forsyth, Giacalone, Glassmeier, Goldstein, Hoeksema,
  Lockwood, Magnes, Maksimovic, Marsch, Matthaeus, Murphy, Nakariakov, Owen,
  Owens, Rodriguez-Pacheco, Richter, Riley, Russell, Schwartz, Vainio, Velli,
  Vennerstrom, Walsh, Wimmer-Schweingruber, Zank, Muller, Zouganelis, \&
  Walsh}]{RN1860}
Horbury, T.~S., O'Brien, H., Blazquez, I.~C., {et~al.} 2020, Astronomy \&
  Astrophysics, 642

\bibitem[{Kasper {et~al.}(2019)Kasper, Bale, Belcher, Berthomier, Case,
  Chandran, Curtis, Gallagher, Gary, Golub, Halekas, Ho, Horbury, Hu, Huang,
  Klein, Korreck, Larson, Livi, Maruca, Lavraud, Louarn, Maksimovic,
  Martinovic, McGinnis, Pogorelov, Richardson, Skoug, Steinberg, Stevens,
  Szabo, Velli, Whittlesey, Wright, Zank, MacDowall, McComas, McNutt, Pulupa,
  Raouafi, \& Schwadron}]{RN1854}
Kasper, J.~C., Bale, S.~D., Belcher, J.~W., {et~al.} 2019, Nature, 576, 228

\bibitem[{Maksimovic {et~al.}(2020)Maksimovic, Bale, Chust, Khotyaintsev,
  Krasnoselskikh, Kretzschmar, Plettemeier, Rucker, Soucek, Steller, Stverak,
  Travnicek, Vaivads, Chaintreuil, Dekkali, Alexandrova, Astier, Barbary,
  Berard, Bonnin, Boughedada, Cecconi, Chapron, Chariet, Collin, de~Conchy,
  Dias, Gueguen, Lamy, Leray, Lion, Malac-Allain, Matteini, Nguyen, Pantellini,
  Parisot, Plasson, Thijs, Vecchio, Fratter, Bellouard, Lorfevre, Danto,
  Julien, Guilhem, Fiachetti, Sanisidro, Laffaye, Gonzalez, Pontet, Queruel,
  Jannet, Fergeau, Brochot, Cassam-Chenai, de~Wit, Timofeeva, Vincent,
  Agrapart, Delory, Turin, Jeandet, Leroy, Pellion, Bouzid, Katra, Piberne,
  Recart, Santolik, Kolmasova, Krupar, Kruparova, Pisa, Uhlir, Lan, Base,
  Ahlen, Andre, Bylander, Cripps, Cully, Eriksson, Jansson, Johansson,
  Karlsson, Puccio, Brinek, Ottacher, Panchenko, Berthomier, Goetz, Hellinger,
  Horbury, Issautier, Kontar, Krucker, Le~Contel, Louarn, Martinovic, Owen,
  {et~al.}}]{RN1861}
Maksimovic, M., Bale, S.~D., Chust, T., {et~al.} 2020, Astronomy \&
  Astrophysics, 642

\bibitem[{Muller {et~al.}(2017)Muller, Nicula, Felix, Verstringe, Bourgoignie,
  Csillaghy, Berghmans, Jiggens, Garcia-Ortiz, Ireland, Zahniy, \&
  Fleck}]{RN1893}
Muller, D., Nicula, B., Felix, S., {et~al.} 2017, Astronomy \& Astrophysics,
  606, 13

\bibitem[{Muller {et~al.}(2020)Muller, St~Cyr, Zouganelis, Gilbert, Marsden,
  Nieves-Chinchilla, Antonucci, Auchere, Berghmans, Horbury, Howard, Krucker,
  Maksimovic, Owen, Rochus, Rodriguez-Pacheco, Romoli, Solanki, Bruno,
  Carlsson, Fludra, Harra, Hassler, Livi, Louarn, Peter, Schuhle, Teriaca, del
  Toro~Iniesta, Wimmer-Schweingruber, Marsch, Velli, De~Groof, Walsh, \&
  Williams}]{RN1884}
Muller, D., St~Cyr, O.~C., Zouganelis, I., {et~al.} 2020, Astronomy \&
  Astrophysics, 642, 31

\bibitem[{Owen {et~al.}(2020)Owen, Bruno, Livi, Louarn, Al~Janabi, Allegrini,
  Amoros, Baruah, Barthe, Berthomier, Bordon, Brockley-Blatt, Brysbaert,
  Capuano, Collier, DeMarco, Fedorov, Ford, Fortunato, Fratter, Galvin,
  Hancock, Heirtzler, Kataria, Kistler, Lepri, Lewis, Loeffler, Marty, Mathon,
  Mayall, Mele, Ogasawara, Orlandi, Pacros, Penou, Persyn, Petiot, Phillips,
  Prech, Raines, Reden, Rouillard, Rousseau, Rubiella, Seran, Spencer, Thomas,
  Trevino, Verscharen, Wurz, Alapide, Amoruso, Andre, Anekallu, Arciuli,
  Arnett, Ascolese, Bancroft, Bland, Brysch, Calvanese, Castronuovo, Cermak,
  Chornay, Clemens, Coker, Collinson, D'Amicis, Dandouras, Darnley, Davies,
  Davison, De~Los~Santos, Devoto, Dirks, Edlund, Fazakerley, Ferris, Frost,
  Fruit, Garat, Genot, Gibson, Gilbert, de~Giosa, Gradone, Hailey, Horbury,
  Hunt, Jacquey, Johnson, Lavraud, Lawrenson, Leblanc, Lockhart, Maksimovic,
  Malpus, Marcucci, Mazelle, {et~al.}}]{RN1863}
Owen, C.~J., Bruno, R., Livi, S., {et~al.} 2020, Astronomy \& Astrophysics, 642

\bibitem[{Owens \& Forsyth(2013)}]{RN1894}
Owens, M.~J. \& Forsyth, R.~J. 2013, Living Reviews in Solar Physics, 10, 50

\bibitem[{Poletto(2015)}]{RN1864}
Poletto, G. 2015, Living Reviews in Solar Physics, 12

\bibitem[{Richardson(2018)}]{RN1886}
Richardson, I.~G. 2018, Living Reviews in Solar Physics, 15, 95

\bibitem[{Rochus {et~al.}(2020)Rochus, Auchere, Berghmans, Harra, Schmutz,
  Schuhle, Addison, Appourchaux, Cuadrado, Baker, Barbay, Bates, BenMoussa,
  Bergmann, Beurthe, Borgo, Bonte, Bouzit, Bradley, Buchel, Buchlin, Buchner,
  Cabe, Cadiergues, Chaigneau, Chares, Cortez, Coker, Condamin, Coumar, Curdt,
  Cutler, Davies, Davison, Defise, Del~Zanna, Delmotte, Delouille, Dolla,
  Dumesnil, Durig, Enge, Francois, Fourmond, Gillis, Giordanengo, Gissot,
  Green, Guerreiro, Guilbaud, Gyo, Haberreiter, Hafiz, Hailey, Halain,
  Hansotte, Hecquet, Heerlein, Hellin, Hemsley, Hermans, Hervier, Hochedez,
  Houbrechts, Ihsan, Jacques, Jerome, Jones, Kahle, Kennedy, Klaproth, Kolleck,
  Koller, Kotsialos, Kraaikamp, Langer, Lawrenson, Le~Clech, Lenaerts, Liebecq,
  Linder, Long, Mampaey, Markiewicz-Innes, Marquet, Marsch, Matthews, Mazy,
  Mazzoli, Meining, Meltchakov, Mercier, Meyer, Monecke, Monfort, Morinaud,
  Moron, Mountney, Muller, Nicula, {et~al.}}]{RN1880}
Rochus, P., Auchere, F., Berghmans, D., {et~al.} 2020, Astronomy \&
  Astrophysics, 642, 21

\bibitem[{Roelof \& Krimigis(1973)}]{RN1883}
Roelof, E.~C. \& Krimigis, S.~M. 1973, Journal of Geophysical Research, 78,
  5375

\bibitem[{{Schatten} {et~al.}(1969){Schatten}, {Wilcox}, \&
  {Ness}}]{1969SoPh....6..442S}
{Schatten}, K.~H., {Wilcox}, J.~M., \& {Ness}, N.~F. 1969, \solphys, 6, 442

\bibitem[{{Schwenn}(1990)}]{1990pihl.book...99S}
{Schwenn}, R. 1990, {Large-Scale Structure of the Interplanetary Medium}, ed.
  R.~{Schwenn} \& E.~{Marsch}, 99

\bibitem[{Schwenn {et~al.}(1978)Schwenn, Montgomery, Rosenbauer, Miggenrieder,
  Muhlhauser, Bame, Feldman, \& Hansen}]{RN1892}
Schwenn, R., Montgomery, M.~D., Rosenbauer, H., {et~al.} 1978, Journal of
  Geophysical Research-Space Physics, 83, 1011

\bibitem[{Solanki {et~al.}(2020)Solanki, del Toro~Iniesta, Woch, Gandorfer,
  Hirzberger, Alvarez-Herrero, Appourchaux, Martinez~Pillet, Perez-Grande,
  Sanchis~Kilders, Schmidt, Cama, Michalik, Deutsch, Fernandez-Rico, Grauf,
  Gizon, Heerlein, Kolleck, Lagg, Meller, Muller, Schuhle, Staub, Albert,
  Copano, Beckmann, Bischoff, Busse, Enge, Frahm, Germerott, Guerrero, Loptien,
  Meierdierks, Oberdorfer, Papagiannaki, Ramanath, Schou, Werner, Yang, Zerr,
  Bergmann, Bochmann, Heinrichs, Meyer, Monecke, Muller, Sperling, Garcia,
  Aparicio, Jimenez, Rubio, Carracosa, Girela, Exposito, Herranz, Labrousse,
  Jimenez, Suarez, Ramos, Barandiaran, Bastide, Campuzano, Cebollero, Davila,
  Fernandez-Medina, Garcia~Parejo, Garranzo-Garcia, Laguna, Martin, Navarro,
  Peral, Royo, Sanchez, Silva-Lopez, Vera, Villanueva, Fourmond, de~Galarreta,
  Bouzit, Hervier, Le~Clec'h, Szwec, Chaigneau, Buttice, Dominguez-Tagle,
  Philippon, Boumier, Le~Cocguen, Baranjuk, Bell, Berkefeld, Baumgartner,
  Heidecke, Maue, Nakai, Scheiffelen, Sigwarth, Soltau, {et~al.}}]{RN1881}
Solanki, S.~K., del Toro~Iniesta, J.~C., Woch, J., {et~al.} 2020, Astronomy \&
  Astrophysics, 642, 35

\bibitem[{Stansby {et~al.}(2020)Stansby, Yeates, \& Badman}]{Stansby2020}
Stansby, D., Yeates, A., \& Badman, S.~T. 2020, Journal of Open Source
  Software, 5, 2732

\bibitem[{Steinvall(2021)}]{Steinvall21}
Steinvall, K. 2021, Astronomy and Astrophysics, submitted

\bibitem[{Szabo {et~al.}(2020)Szabo, Larson, Whittlesey, Stevens, Lavraud,
  Phan, Wallace, Jones-Mecholsky, Arge, Badman, Odstrcil, Pogorelov, Kim,
  Riley, Henney, Bale, Bonnell, Case, de~Wit, Goetz, Harvey, Kasper, Korreck,
  Koval, Livi, MacDowall, Malaspina, \& Pulupa}]{RN1877}
Szabo, A., Larson, D., Whittlesey, P., {et~al.} 2020, Astrophysical Journal
  Supplement Series, 246

\bibitem[{Thieme {et~al.}(1990)Thieme, Marsch, \& Schwenn}]{RN926}
Thieme, K.~M., Marsch, E., \& Schwenn, R. 1990, Annales Geophysicae-Atmospheres
  Hydrospheres and Space Sciences, 8, 713

\bibitem[{Viall \& Borovsky(2020)}]{RN1891}
Viall, N.~M. \& Borovsky, J.~E. 2020, Journal of Geophysical Research-Space
  Physics, 125, 35

\bibitem[{Wang \& Sheeley(1990)}]{RN1890}
Wang, Y.~M. \& Sheeley, N.~R. 1990, Astrophysical Journal, 355, 726

\end{thebibliography}


\end{document}